\documentclass[twocolumn,prd,longbibliography,
 amsmath,amssymb,superscriptaddress]{revtex4-2}
\usepackage{graphicx}
\usepackage{dcolumn}
\usepackage{bm}
\usepackage[colorlinks=true, citecolor=blue, linkcolor=blue, urlcolor=blue]{hyperref}
\usepackage{xcolor}
\usepackage{amsmath, amssymb, amsfonts}
\usepackage{mathtools}
\usepackage{subfigure}
\usepackage{mathrsfs}
\usepackage{sidecap}
\usepackage{verbatim}

\definecolor{purple1}{rgb}{128,0,128}

\usepackage{bigints}
\newcommand{\bea}{\begin{eqnarray}}
\newcommand{\ea}{\end{eqnarray}}

\definecolor{darkpastelgreen}{rgb}{0.01, 0.75, 0.24}

\def\sgn{\mbox{sgn}}

\def\d{\mathrm{d}}
\newcommand{\floor}[1]{\left\lfloor #1 \right\rfloor}

\usepackage{diagbox}
\usepackage{float}       
\usepackage{siunitx}
\usepackage{booktabs} 
\begin{document}

\title{
Solving the nonlinear Klein-Gordon equation: semianalytical Galerkin method} 
\author{Annibal D. \surname{de Figueiredo Neto}}
\email{annibalfig@unb.br}
\affiliation{International Center of Physics, Institute of Physics, University of Brasilia, 70297-400 Brasilia, Federal District, Brazil} 
\author{Caio C. \surname{Holanda Ribeiro}}
\email{caiocesarribeiro@alumni.usp.br}
\affiliation{International Center of Physics, Institute of Physics, University of Brasilia, 70297-400 Brasilia, Federal District, Brazil} 
\author{Luana L. \surname{Silva Ribeiro}}
\email{luana.ribeiro@unifei.edu.br}
\affiliation{Instituto de Ciências Puras e Aplicadas, Universidade Federal de Itajubá, 35903-087 Itabira, Minas Gerais, Brazil} 

\date\today

\begin{abstract}
In this work, approximate solutions to the nonlinear Klein-Gordon equation are constructed by means of the Galerkin method. Specifically, it is shown how the dynamics of a real scalar field in $1+1$ dimensions subjected to Dirichlet boundary conditions and Mexican-hat-like potentials can be approximated by mechanical systems with a few particles. Because the approximation is performed at a Lagrangian level, one of the advantages of this method is the control over conservation laws present in the field theory, which are captured by the finite mechanical systems. Among the results, exact stationary solutions for the nonlinear KG equation are found in terms of Jacobi elliptic functions, which are shown to correspond to stationary configurations of the mechanical systems. Furthermore, numerical simulations are provided, giving hints towards the convergence of the method.

\end{abstract}

\maketitle

\section{Introduction}

The Standard Model of Particle Physics is a theory with one of the best predictive powers ever created and it is believed to be the correct framework for studying all elementary process not involving gravity \cite{RevModPhys.71.S96}. Just to quote a few examples, we cite the Higgs mechanism \cite{Higgs,ObservationofHiggsBoson}, in which a fundamental field, the Higgs field, provides an effective mass for some of Nature's gauge bosons and the prediction of the electron anomalous magnetic moment, that agrees with experiments to $1$ part in $10^{12}$ \cite{Fan2023}. The key idea upon which the Standard Model is based is that fields act as platforms to study fundamental particles and their interactions, and this interpretation emerges naturally when the (free) fields are expanded in suitable eigenfunction bases, e.g., plane waves in Minkowski spacetime \cite{peskin}.

In general, however, exact solutions for quantum field models in nonlinear theories are either not possible to find or only exist for particular scenarios \cite{PhysRevLett.121.110402}. Such problems can be circumvented, for instance, if perturbative analyses are possible \cite{peskin}, which greatly restrict the field models that can be studied. In this context, an interesting approach consists in studying ``semianalytical'' methods, in which approximations are built not for the solutions, but for the field dynamics \cite{10.21468/SciPostPhysCore.6.1.003}. This is the key idea, for instance, of Galerkin (spectral) methods \cite{fletcher1984computational}, that were introduced in 1915 in the study of elastic structures \cite{galerkin1915series}. Since then, the method has been applied to solve ordinary, partial, and integro-differential equations \cite{fletcher1984computational}. In particular, this method is the base for the construction of the Lorenz model in the context of fluid dynamics \cite{DeterministicNonperiodicFlow}.

Using as motivation the Higgs field, in this article we consider the Galerkin method to find approximate solutions to the nonlinear Klein-Gordon equation for a {\it real} scalar field in $1+1$ dimensions. Specifically, we consider a scalar field subjected to Mexican-hat-like potentials, which are associated to the Higgs mechanism \cite{Higgs}. Also, the field is subjected to external potentials as to ensure Dirichlet boundary conditions. Following the same philosophy of particle physics, the method is inspired from a description of the field dynamics in terms of mechanical systems of infinitely many particles, whose properties depend on the particular basis adopted to expand the field variable. By considering a family of mechanical systems with a finite number of particles, we then show how to construct approximate solutions to the different regimes of the nonlinear Klein-Gordon equation. Our results include the construction of exact stationary solutions as function of the potential parameters, and how these are connected to the stationary solutions of the mechanical systems. Furthermore, it is shown, via numerical simulations, how these mechanical systems capture the dynamics of the field theory, pointing towards the convergence of the method.

Our work is organized as follows. Section \ref{lagrangianformulation} presents the Lagrangian formulation for the field theory, and its equivalent description in terms of a mechanical system with an infinite number of particles. In Section \ref{stationaryKG} we construct stationary solutions to the nonlinear Klein-Gordon equation in terms of Jacobi elliptic functions. Section \ref{approxfinite} presents the finite mechanical systems, and how they approximate the field theory. The work is concluded with final remarks in Section \ref{finalremarks}.

\section{The Lagrangian formulation}
\label{lagrangianformulation}

We consider a real scalar field $\varphi=\varphi(t,x)$ in $1+1$ dimensions. By employing the usual notation ($c=1$) $x^{\mu}=(t,x)$, and metric $\eta_{\mu\nu}=\mbox{diag}(1,-1)$, we assume that the theory is ruled by the Lagrangian density  
\begin{equation}
    \mathcal{L}=\frac{1}{2}(\partial_\mu\varphi)(\partial^\mu\varphi)-V(\varphi),\label{lagrangian}
\end{equation}
where the Einstein summation convention is adopted. Note that because $V$ does not have explicit dependence on $x^{\mu}$, the canonical stress tensor
\begin{equation}
    T^{\mu}_{\ \nu}=\frac{\partial\mathcal{L}}{\partial(\partial_\mu\varphi)}\partial_\nu \varphi-\delta^{\mu}_{\ \nu}\mathcal{L},
\end{equation}
is such that $\partial_\mu T^{\mu}_{\ \nu}=0$. This continuity equation in turn implies that the system Hamiltonian $H=\int\d x \mathcal{H}$, written in terms of the Hamiltonian density $\mathcal{H}=T^{0}_{\ 0}$, i.e.,
\begin{align}
    H&=\int\d x\left[\frac{1}{2}(\partial_t\varphi)^2+\frac{1}{2}(\partial_x\varphi)^2+V\right],\label{hamiltonian1}
\end{align}
is a conserved constant when the net energy flux entering the system vanishes. In particular, this is always the case if the energy flux, $T^{1}_{\ 0}$, vanishes at the spatial boundary. We note also that the continuity equation leads to a conservation law for the system momentum, $P=-\int\d x T^{0}_{\ 1}$. However, in this work we are interested in the case in which the scalar field is confined to a compact spatial region by external laboratory agents. Thus, forces are exerted on the system, and its momentum will not be a conserved quantity   in general.

We are interested in the specific non-linear theory such that 
\begin{align}
    V(\varphi)=\frac{\beta}{4}(\varphi^2-\varphi_0^2)^2.\label{potentialV}
\end{align}
Here, the constant $\beta$, which we allow to assume any real value, measures the intensity of the nonlinear effects.
\begin{figure}
    \centering
    \includegraphics[width=1\linewidth]{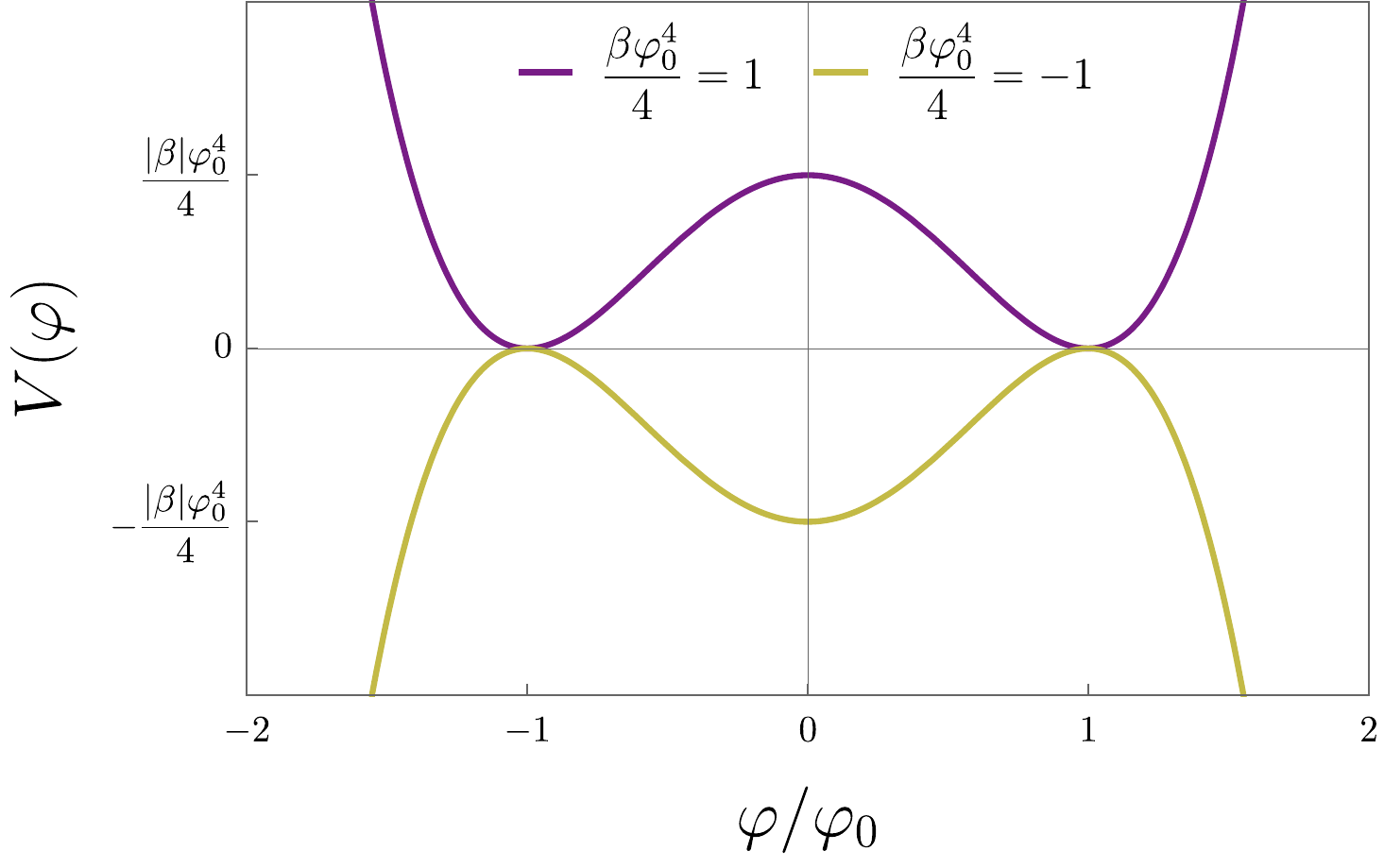}
    \caption{The ``1D'' Mexican-hat and inverted Mexican-hat potentials. Note that for negative $\beta$ the energy density can assume negative values.}
    \label{figpot}
\end{figure}
Figure \ref{figpot} shows the behavior of $V(\varphi)$ for a couple of parameter choices. Notice that if $\beta>0$ one obtains the typical behavior of a Mexican-hat potential, and the corresponding Hamiltonian \eqref{hamiltonian1} is a positive definite functional, whereas is $\beta<0$, a sort of inverted Mexican-hat potential is obtained, with an unbounded (from below) Hamiltonian. In this work, as we are interested in solving the classical field equation, we consider both cases, as runaway solutions are of no concern. 

The Euler-Lagrange equation for this theory is the nonlinear Klein-Gordon (KG) equation
\begin{equation}
    \frac{\partial^2\varphi}{\partial t^2}-\frac{\partial^2\varphi}{\partial x^2}+\beta(\varphi^2-\varphi_0^2)\varphi=0.\label{kgequation}
\end{equation}
We restrict our analysis to the case where the system is confined to the interval $[0,\ell]$, and we assume that $\varphi$ satisfies Dirichlet boundary conditions (BC) on $x=0,\ell$. In this case, the energy flux vanishes at the boundaries and $H$ is constant.
%
%

\subsection{Formulation in terms of an equivalent mechanical system}

The Galerkin method consists in associating to the theory described by Eq.~\eqref{lagrangian}  a system of infinitely many interacting particles as follows. Let $\{\phi_{n}\}_n$ be a complete set of {\it real} eigenfunctions  of the operator $-\partial_x^2$ on the interval $[0,\ell]$ \cite{Mohammed2008} subjected to the same boundary conditions as $\varphi$:
\begin{equation}
  -\partial_x^2\phi_n=\kappa_n^2\phi_n,  
\end{equation}
with 
\begin{equation}
    \kappa_n=\frac{n\pi}{\ell}.
\end{equation}
The eigenfunctions for Dirichlet BC read
\begin{equation}
    \phi_n(x)=\mathcal{N}_{n}\sin\left(\kappa_nx\right), \ \  n=1,2,3,\ldots,\label{dirichletbf}
\end{equation}
%
%
%
where $\mathcal{N}_n=\sqrt{2/\ell}$ if $n>0$. We also define $\mathcal{N}_0=1/\sqrt{\ell}$, that will be important later on. The $\phi_n$ are normalized according to the standard inner product
\begin{equation}
    \langle\phi_n,\phi_m\rangle\equiv\int_0^\ell\d x\phi_n\phi_m^*=\delta_{nm}.
\end{equation}
Therefore, we can expand $\varphi$ in terms of $\phi_n$ as
\begin{equation}
    \varphi(t,x)=\sum_nB_n(t)\phi_n(x).\label{fourierexpansion}
\end{equation}
The Fourier coefficients, $B_n(t)$, are fixed by the orthogonality of the $\phi_n$:
\begin{equation}
    B_n=\langle\varphi,\phi_n\rangle.\label{fouriercoefficients}
\end{equation}
With the expansion \eqref{fourierexpansion}, the Lagrangian $L=\int_0^\ell\d x\mathcal{L}$ becomes
\begin{align}
    L=&\frac{1}{2}\sum_n\left[\left(\frac{\d B_n}{\d t}\right)^2-\left(\kappa_n^2-\beta\varphi_0^2\right)B_n^2\right]\nonumber\\
    &-\frac{\beta}{4\ell}\sum_{nmpq}D_{nmpq}B_nB_mB_pB_q-\ell\frac{\beta\varphi_0^4}{4},\label{lagrangianreduced}
\end{align}
where
\begin{equation}
    D_{nmpq}=\ell\int_0^\ell\d x\phi_n\phi_m\phi_p\phi_q,
\end{equation}
is a (dimensionless)  tensor of coupling constants. Note that $D_{nmpq}$ is symmetric (because the eigenfunctions are real) under any permutation of its indices. 
With the aid of the eigenfunctions, we find that
\begin{align}
    D_{nmpq}=&\frac{1}{\ell \mathcal{N}_{n-m}^2}(\delta_{|n-m|,|p-q|}-\delta_{|n-m|,p+q})\nonumber\\
    &-\frac{1}{\ell \mathcal{N}_{n+m}^2}(\delta_{n+m,|p-q|}-\delta_{n+m,p+q}).
    \end{align}
%
%

Here an interesting observation is in order. Equation \eqref{lagrangianreduced} is the Lagrangian for a system of infinitely many particles, one for each $n$. Each such particle has position given by $B_n=B_n(t)$ and it is subjected to a harmonic potential with squared frequency $\kappa_n^2-\beta\varphi_0^2$. Furthermore, if $\beta\neq0$, these particles interact with coupling constant modulated by the tensor $D_{nmpq}$. This is, in fact, the core idea behind the standard model of particle physics, in which fields become platforms to describe particle physics after  suitable bases of eigenfunctions are fixed. It is instructive to work with normalized variables by defining $B_n=a A_n$ and $t=\ell \tau/\pi$, where 
\begin{align}
    a=\frac{\pi}{\sqrt{\ell|\beta|}}.\label{adef}
\end{align}
With this prescription, the Lagrangian becomes
\begin{align}
    &\frac{\ell^2}{\pi^2a^2}L=\frac{1}{2}\sum_n\left[\dot{A}_n^2-(n^2+\lambda)A_n^2\right]\nonumber\\
    &+\frac{\sgn(\lambda)}{4}\sum_{nmpq}D_{nmpq}A_nA_mA_pA_q+\frac{\sgn(\lambda)}{4}\lambda^2,\label{lagrangianreduced2}
\end{align}
with
\begin{equation}
\lambda=-\beta \varphi_0^2\frac{\ell^2}{\pi^2},    
\end{equation} 
being the only free parameter of the theory. Also, $\dot{A}_n=\d A_n/\d\tau$. Equation \eqref{lagrangianreduced2} is the basic Lagrangian for our analysis. The Euler-Lagrange equation for $A_n$ then reads
\begin{equation}
    \ddot{A}_n+(n^2+\lambda)A_n=\sgn(\lambda)\sum_{mpq}D_{nmpq}A_mA_pA_q,\label{infsystem}
\end{equation}
where $n=1,2,\ldots$ and it is, by construction, equivalent to the nonlinear KG equation.

Note also that the Hamiltonian $H$, Eq.~\eqref{hamiltonian1}, becomes
\begin{align}
    &\frac{\ell^2}{\pi^2a^2}H=\frac{1}{2}\sum_n\left[\dot{A}_n^2+(n^2+\lambda)A_n^2\right]\nonumber\\
    &-\frac{\sgn(\lambda)}{4}\sum_{nmpq}D_{nmpq}A_nA_mA_pA_q-\frac{\sgn(\lambda)}{4}\lambda^2,\label{hamiltonian}
\end{align}
and it coincides with the Hamiltonian obtained directly from the Lagrangian \eqref{lagrangianreduced2} by treating $A_n$ as generalized coordinates. Furthermore, the equation of motion also follows from a potential $U$: $\ddot{A}_n=-\partial U/\partial A_n$, where
\begin{align}
    U=&\frac{1}{2}\sum_n(n^2+\lambda)A_n^2-\frac{\sgn(\lambda)}{4}\sum_{nmpq}D_{nmpq}A_nA_mA_pA_q\nonumber\\
    &-\frac{\sgn(\lambda)}{4}\lambda^2.\label{potentialfull}
\end{align}
Note that stationary solutions are precisely the ones given by the critical points of $U$.

\section{Stationary solutions for the Klein-Gordon equation}
\label{stationaryKG}

Usually, one needs to solve the KG equation in two types of problems, namely, in Cauchy problems and in boundary value problems. In the latter, one seeks solutions that satisfy the boundary conditions, whereas in the former, one starts from a suitable initial condition, $\varphi(t,x)|_{t=t_0}$, $\partial_t\varphi(t,x)|_{t=t_0}$, and aims to find $\varphi(t,x)$ for $t>t_0$. In both cases, time-independent solutions might exist, which is a consequence of the nonlinear term added to the KG equation. In fact, recall that when $\lambda$ (or $\beta$) vanishes, the only stationary solution of the KG equation subjected to Dirichlet BC is the trivial solution. In this section we present exact stationary solutions for the KG equation for any value of $\lambda$. 

If $\varphi(t,x)\equiv a f(x)$ [cf.~Eq.~\eqref{adef}] is a stationary solution of the nonlinear KG equation, then
\begin{equation}
    \frac{\d^2 f}{\d x^2}-\frac{\pi^2\lambda}{\ell^2} f+\frac{\pi^2}{\ell} \sgn(\lambda)f^3=0,\label{odetosolve}
\end{equation}
with $f(x)$ subjected to Dirichlet BC. Real solutions of the above equation can be found in terms of Jacobi elliptic functions $\mbox{sn}(u,k)$ and $\mbox{cn}(u,k)$,  defined implicitly by \cite{specialfun-wang}
\begin{align}
u&=\int_0^{\mbox{sn}(u,k)}\frac{\d t}{\sqrt{(1-t^2)(1-k^2t^2)}},\label{sndef}\\
    u&=\int_1^{\mbox{cn}(u,k)}\frac{\d t}{\sqrt{(1-t^2)(k^{'2}-k^2t^2)}},\label{cndef}
\end{align}
with $k'=\sqrt{1-k^2}$. These special functions appear, for instance, in the study of the quartic oscillator \cite{lawden1989elliptic}.

The functions $\mbox{sn}(u,k)$ and $\mbox{cn}(u,k)$ possess a number of useful properties. For our purposes, we note that the zeros of both $\mbox{sn}(u,k)$ and $\mbox{cn}(u,k)$ are known:
\begin{align}
    \mbox{sn}(u,k)&=0\Rightarrow u=2nK(k)+2n'iK(k'),\\
    \mbox{cn}(u,k)&=0\Rightarrow u=(2n+1)K(k)+2n'iK(k'),
\end{align}
where $K(k)$ is the Complete Elliptic Integral \cite{Gradius}, and $n,n'$ are arbitrary integers. It is instructive to study the cases in which $\lambda$ is either positive or negative separately.

\subsection{Negative $\lambda$}

\begin{figure}
    \includegraphics[width=1\linewidth]{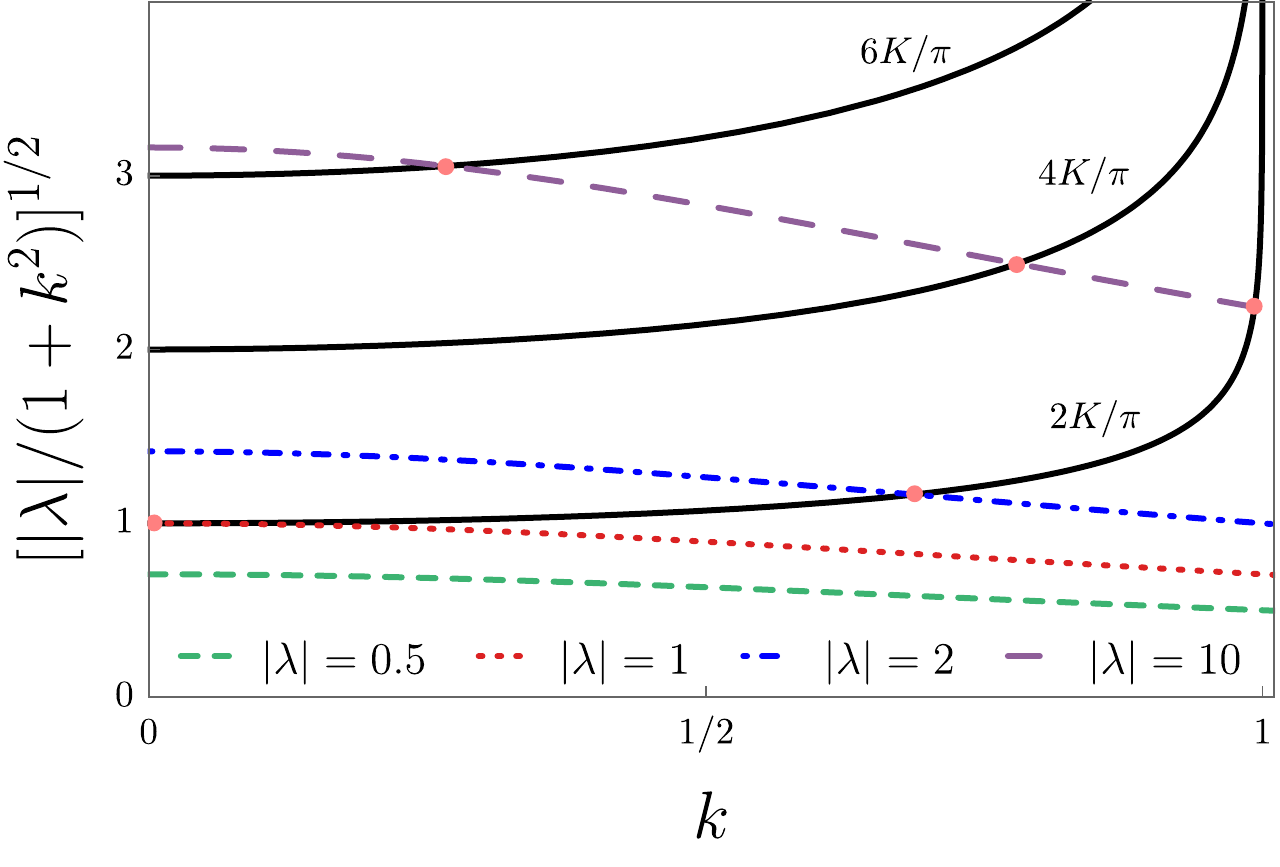}
    \caption{Graphical determination of the values of $k_4$ for the case of negative $\lambda$. The continuous curves are the positive multiples of $2K/\pi$, whereas the short-dashed, dotted, dot-dashed, and long-dashed curves are the function $[|\lambda|/(1+k^2)]^{1/2}$ for $-\lambda=0.5, 1, 2$, and $10$, respectively. The light red dots where both functions meet are the solutions for $k_4$. Note that no solutions other than the trivial one exist for $|\lambda|\leq1$.}
    \label{figsn}
\end{figure}

\begin{figure}[H]
    \centering
    \includegraphics[width=1\linewidth]{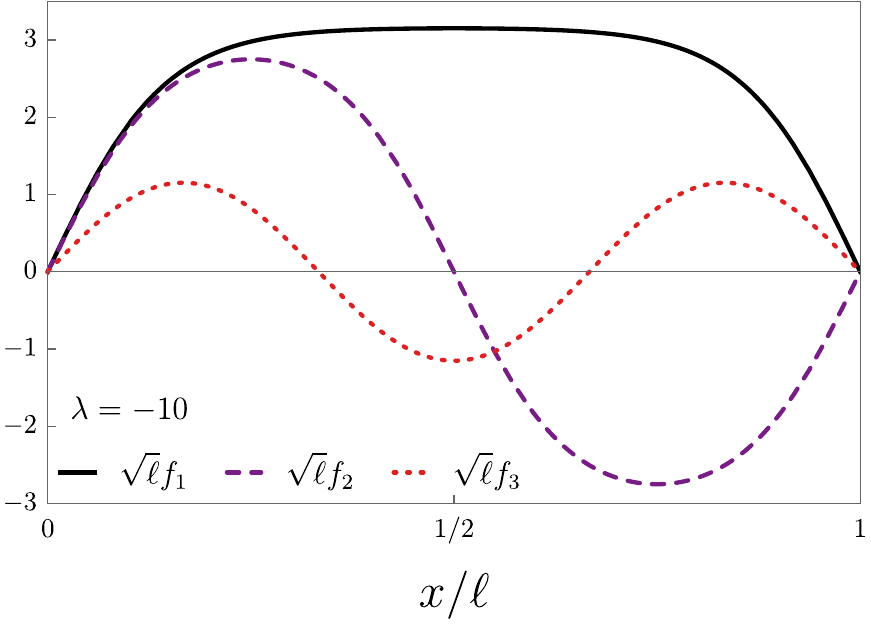}
    \caption{The first three stationary solutions for the nonlinear KG equation, according to their energy, when $\lambda=-10$. Here, $f_1$ corresponds to $k_4=0.993$, $f_2$ to $k_4=0.780$, and $f_3$ to $k_4=0.267$.}
    \label{figsnexample}
\end{figure}

When $\lambda<0$, solutions can be found in the form
\begin{equation}
    f(x)=\sqrt{\frac{2}{\ell}}k_1\mbox{sn}(k_2 \pi x/\ell+k_3,k_4),
\end{equation}
where $k_i$, $i=1,2,3,4$, are real parameters. A similar ansatz was used to build traveling solutions in a different context in \cite{Ates2017}.  It is straightforward to show, from Eq.~\eqref{sndef}, that
\begin{align}
    \frac{\d^2\ }{\d u^2}\mbox{sn}(u,k)&=-(1+k^2)\mbox{sn}(u,k)+2k^2\mbox{sn}^3(u,k).
\end{align}
With this equation, we can show that
\begin{align}
   k_2^2&=|\lambda|\frac{1}{1+k_4^2},\\
    k_1^2&=|\lambda|\frac{k_4^2}{1+k_4^2},
\end{align}
with $k_3$ and $k_4$ arbitrary integration constants. These can be fixed by specifying the value of the single parameter $\lambda$. For Dirichlet BC we find that $k_2,k_3$ are solutions of
\begin{equation}
    \mbox{sn}(k_3,k_4)=\mbox{sn}(\pi k_2+k_3,k_4)=0,
\end{equation}
which imply that
\begin{align}
    k_3&=2n'K(k_4),\label{conditionsn2}\\
    k_2&= \left(\frac{|\lambda|}{1+k_4^2}\right)^{1/2}=\frac{2nK(k_4)}{\pi}.\label{conditionsn}
\end{align}
We fixed a positive value for $k_2$ in terms of $k_4$ in equation \eqref{conditionsn}, which in turn determines the possible values of $k_4$. Figure \ref{figsn} shows the functional behavior of $2K/\pi$ and $[|\lambda|/(1+k^2)]^{1/2}$ as function of $k$ for several values of $\lambda$.
The coincidences between the function $[|\lambda|/(1+k^2)]^{1/2}$ and positive multiples of $2K/\pi$ give rise to admissible stationary solutions. Also, if $k_4=0$ is a coincidence, the corresponding solution is identically zero, and thus only positive $k_4$ produce relevant solutions. Note that $2K/\pi$ is increasing in $[0,1)$, tends to $1$ when $k\rightarrow0$, and diverges as $k\rightarrow1$. Also, $[|\lambda|/(1+k^2)]^{1/2}$ is decreasing, and thus nontrivial solutions exist only when $|\lambda|>1$ [see Fig.~\ref{figsn}]. The number of distinct solutions is $\floor{|\lambda|^{1/2}}$ when $|\lambda|^{1/2}$ is not an integer, and $\floor{|\lambda|^{1/2}}-1$ otherwise. Here, $\floor{x}$ is the floor function. Finally, we observe that once $k_4$ is fixed, the remaining parameter $k_3$ is fixed up to an integer by Eq.~\eqref{conditionsn2}. This, however, will only change the overall sign of the solution due to the symmetry properties of $\mbox{sn}(u,k)$ \cite{Gradius}, and thus we can take $k_3=0$ without loss of generality. When there are more than one stationary solution for a given $\lambda$, we will label then according to the absolute value of the energy they store, calculated according to Eq.~\eqref{hamiltonian1}. For instance, $f_1$ will denote the stationary solution corresponding to the smallest energy value and so on. Figure \ref{figsnexample} shows the first three solutions for $\lambda=-10$.

Finally, we recall that, in view of the expansion \eqref{fourierexpansion}, stationary solutions correspond to time-independent Fourier coefficients $B_n$, or, equivalently, $A_n$. Let 
\begin{equation}
    f_i(x)=\sum_{n=1}^{\infty}\mathcal{A}_{i,n}\phi_n(x),
\end{equation}
such that $\mathcal{A}_{i,n}=\langle f_i,\phi_n\rangle$. We present in Table \ref{tab1-An} the values of the first Fourier coefficients for the solutions shown in Fig.~\ref{figsnexample}. In what follows, we show how these coefficients can be approximated by considering systems with only a few particles.
\begin{table}[H]
\centering
\begin{tabular}{c|S  S S}
\diagbox[width=2.8em,height=1.5em]{$n$}{$i$} 
  & \multicolumn{1}{c}{1} 
  & \multicolumn{1}{c}{2} 
  & \multicolumn{1}{c}{3} \\
\toprule
\hline
1 &  2.62567     & 0   & 0. \\
2 & 0 & 2.05109   & 0 \\
3 &  0.493473    &0   & 0.818358 \\
4 & 0 & 0   & 0 \\
5 &  0.119402    & 0   & 0 \\
6 & 0 & 0.112618   & 0 \\
7 &  0.0292736   & 0   & 0 \\
8 &  0   & 0   & 0 \\
9 &  0.00718   & 0   & 0.00375 \\
10 &  0   & 0.00656   & 0 \\
\hline
$\frac{\ell^2H}{\pi^2a^2}$ & 9.49008 &18.77293 &24.83282\\
\bottomrule
\end{tabular}
\caption{Several values of $\mathcal{A}_{i,n}$, with $5$ precision digits, of the first three stationary solutions for the case $\lambda=-10$. Here we show only the first 10 coefficients.}
\label{tab1-An}
\end{table}

\subsection{Positive $\lambda$}

\begin{figure}
    \centering
    \includegraphics[width=1\linewidth]{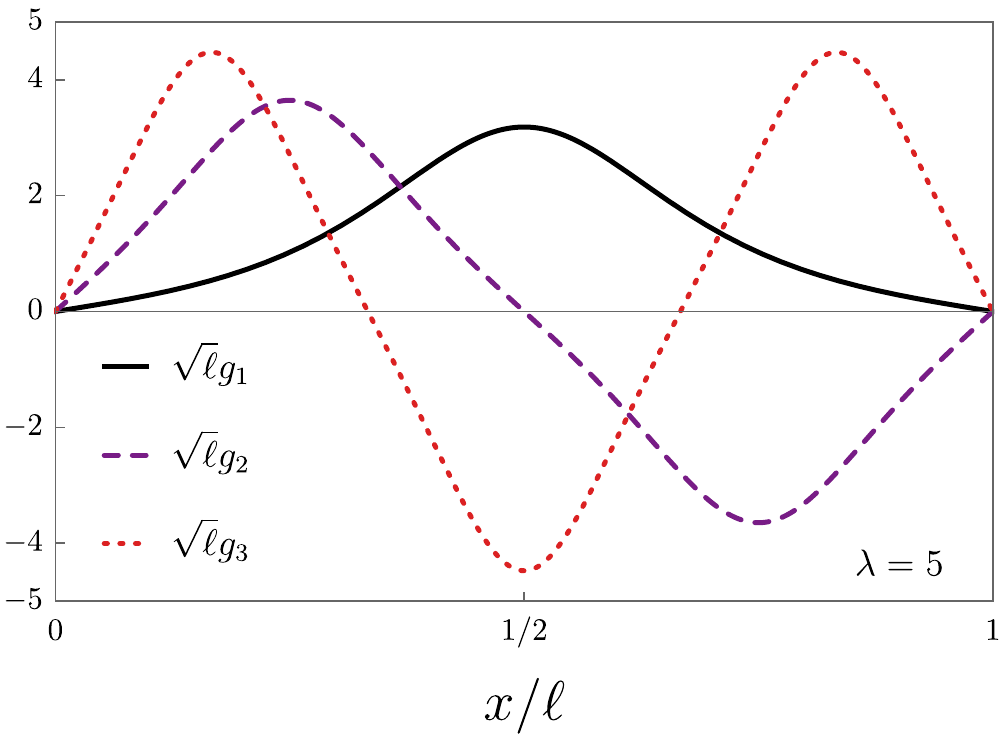}
    \caption{The first three stationary solutions, according to the energy they store, for the case of $\lambda=5$. Here, $g_1$ corresponds to the solution with $q_4=0.87$, $g_2$ to $q_4=0.76$, and $g_3$ to $q_4=0.73$.}
    \label{figlambda5}
\end{figure}

The case of positive $\lambda$ is peculiar, for the Hamiltonian is not bounded from below. The first difference in solving the nonlinear equation \eqref{odetosolve} in comparison to the previous case is that now solutions are described by $\mbox{cn}$ instead of $\mbox{sn}$. Let us take
\begin{equation}
    g(x)=\sqrt{\frac{2}{\ell}}q_1\mbox{cn}\left(q_2 \pi x/\ell+q_3,q_4\right),
\end{equation}
where $q_i$, $i=1,2,3,4$, are real parameters. 
%
%
It follows from Eq.~\eqref{cndef} that
\begin{align}
    \frac{\d^2\ }{\d u^2}\mbox{cn}(u,k)&=(2k^2-1)\mbox{cn}(u,k)-2k^2\mbox{cn}^3(u,k),
\end{align}
and the only constraints on the constants are
\begin{align}
   q_2^2&=\frac{|\lambda|}{2q_4^2-1},\\
    q_1^2&=q_2^2q_4^2.
\end{align}
Also, by imposing that $q_2$ must be real, we find that
\begin{align}
 \frac{1}{2}<q_4^2<1.  
\end{align}
The rightmost inequality is necessary to guarantee that $\mbox{cn}(u,q_4)$ is  real for all $u$  \cite{Gradius}.
The constants $q_i$ can be fixed by specifying the value of the parameter $\lambda$. For Dirichlet BC we find that
\begin{equation}
    \mbox{cn}(q_3,q_4)=\mbox{cn}\left(\pi q_2+q_3,q_4\right)=0,\label{dirichletbc}
\end{equation}
implying
\begin{align}
    q_3&=(2n'+1)K(q_4),\label{conditioncn2}\\
     q_2&= \left(\frac{|\lambda|}{2q_4^2-1}\right)^{1/2}=\frac{2nK(q_4)}{\pi},\label{conditioncn1}
\end{align}
for integers $n,n'$. In the same fashion as occurs for the negative $\lambda$ case, Eq.~\eqref{conditioncn1} determines the possible values for $q_4$, which in turn fixes the corresponding $q_3$ by means of Eq.~\eqref{conditioncn2} up to an odd integer multiple. Different values of $q_3$ will produce, however, the same solutions up to an overall sign due to the symmetries of $\mbox{cn}(u,k)$ \cite{Gradius}, and thus we can take $q_3=K(q_4)$ without loss of generality. 
\begin{figure}
    \centering
    \includegraphics[width=1\linewidth]{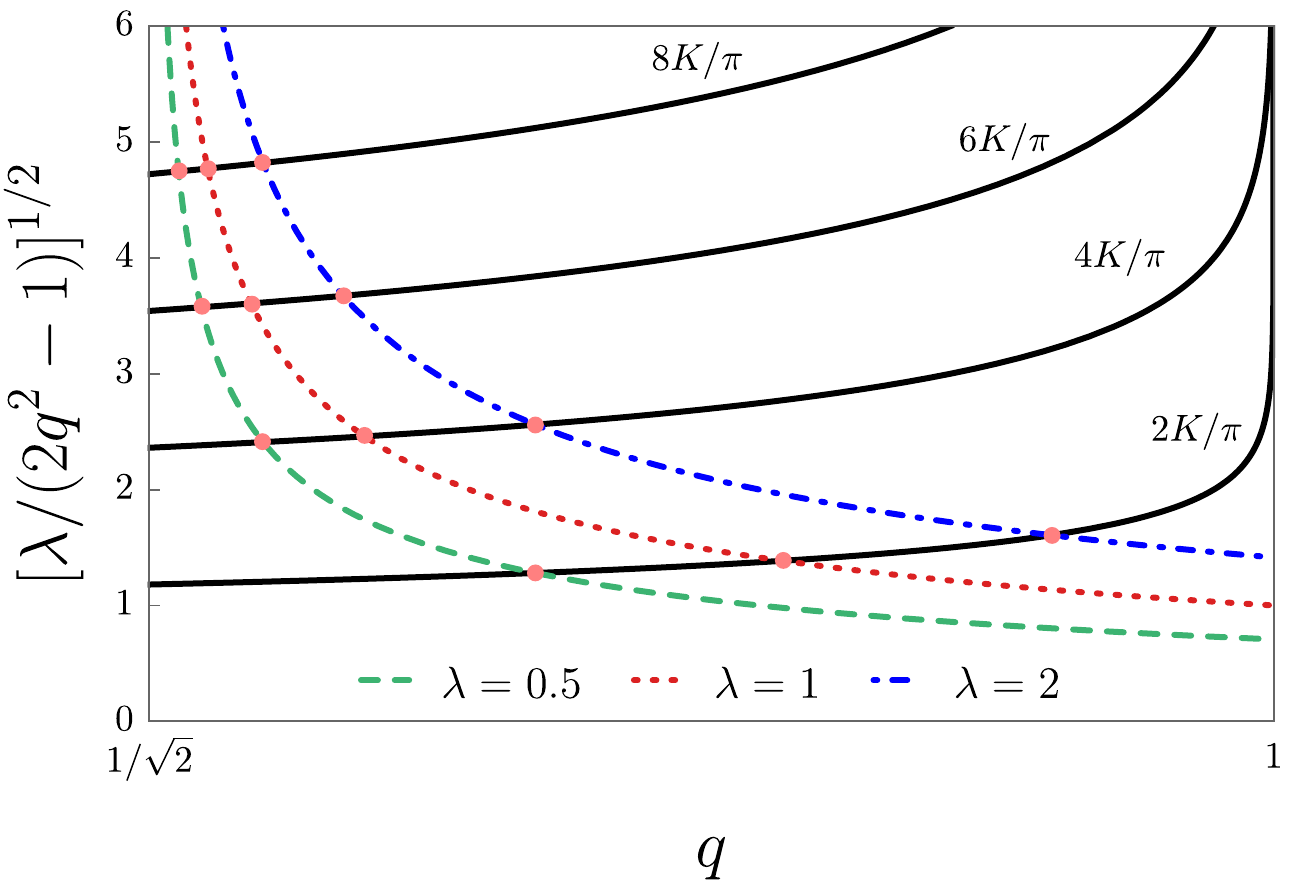}
    \caption{Representation of the solutions for $q_4$ found by means of Eq.~\eqref{conditioncn1}. Notice that, differently from the negative $\lambda$ case, the divergence of $[\lambda/(2q^2-1)]$ when $q\rightarrow1/\sqrt{2}$ ensures the existence of infinitely many nontrivial solutions (light red dots) for $q_4$ for any value of $\lambda>0$.}
    \label{figlambdapos}
\end{figure}
Figure \ref{figlambdapos} depicts the coincidences given by Eq.~\eqref{conditioncn1}. In striking difference from the previous case, when $\lambda>0$, because the function $[\lambda/(2q^2-1)]$ tends to $\infty$ when $q\rightarrow1/\sqrt{2}$ from above, for each value of $\lambda>0$ there exists an infinite number of distinct solutions. Figure \ref{figlambda5} shows the first three solutions for the case $\lambda=5$. We use the same labeling as in the previous case, and let $g_1,g_2,\ldots$ denote the distinct solutions with increasing absolute energy value.
%
Furthermore, Table \ref{tab1-Cn} presents some of the stationary Fourier coefficients, following the notation
\begin{equation}
    g_i(x)=\sum_{n=1}^{\infty}\mathcal{C}_{i,n}\phi_n(x).
\end{equation}
\begin{table}[H]
\centering
\begin{tabular}{c|S  S S}
\diagbox[width=2.8em,height=1.5em]{$n$}{$i$} 
  & \multicolumn{1}{c}{1} 
  & \multicolumn{1}{c}{2} 
  & \multicolumn{1}{c}{3} \\
\toprule
\hline
1 &  1.59777   & 0           & 0. \\
2 & 0          & 2.29772     & 0 \\
3 & -0.488998  &0            & 2.93395 \\
4 & 0          & 0           & 0 \\
5 & 0.123441   & 0           & 0 \\
6 & 0          & -0.251878   & 0 \\
7 & -0.0307449 & 0           & 0 \\
8 & 0          & 0           & 0 \\
9 & 0.00765108 & 0           & -0.213711 \\
10 & 0         & 0.0251559   & 0 \\
\hline
$\frac{\ell^2H}{\pi^2a^2}$ & -1.45534 &6.29612 &24.8724\\
\bottomrule
\end{tabular}
\caption{$\mathcal{C}_{i,n}$ values  with $5$ precision digits for the case of $\lambda=5$ for the first three solutions. Here we show only the first 10 coefficients.}
\label{tab1-Cn}
\end{table}


\section{Approximations using mechanical systems with a finite number of particles}
\label{approxfinite}

The Lagrangian \eqref{lagrangianreduced2} describes a system of infinitely many interacting particles, each of them with position given by $A_{n}=A_{n}(t)$. This suggests considering, for each integer $N\geq1$, a mechanical system composed by $N$ interacting particles whose positions are given by $A_{n}^{(N)}=A_{n}^{(N)}(t)$, $1\leq n\leq N$, with Lagrangian (up to an overall constant)
\begin{align}
    &L^{(N)}=\frac{1}{2}\sum_{n=1}^{N}\left(\dot{A}^{(N)}_n\right)^2-U^{(N)},
    \label{lagrangiantruncated}
\end{align}
where
\begin{align}
    U^{(N)}=&\frac{1}{2}\sum_{n=1}^{N}(n^2+\lambda)\left(A^{(N)}_{n}\right)^2\nonumber\\
    &-\frac{\sgn(\lambda)}{4}\sum_{nmpq}D_{nmpq}A^{(N)}_nA^{(N)}_mA^{(N)}_pA^{(N)}_q\nonumber\\
    &-\frac{\sgn(\lambda)}{4}\lambda^2.
\end{align}
Also, we set $A^{(N)}_{n}=0$ for $n>N$. The reasons for considering such a system are twofold. First, note that for ``sufficiently'' well-behaved solutions of the KG equation, the Fourier coefficients $B_{n}$ of Eq.~\eqref{fourierexpansion}, or, equivalently, the $A_n$, go to zero as $n\rightarrow \infty$. Therefore we expect that smooth solutions of the KG equation can be well approximated by a mechanical system described by Eq.~\eqref{lagrangiantruncated} for a sufficiently large $N$. Note that this approximation might not hold when shock or other types of waves exist \cite{PAIVA20211196}. Second, the system described by Eq.~\eqref{lagrangiantruncated}, which is inspired by the underlying dynamics of an interacting field, has the potential of displaying interesting physical properties, like chaos in the example of the Lorenz system \cite{FiniteAmplitudeFreeConvectionasanInitialValueProblemI,DeterministicNonperiodicFlow}. These can be used to infer information about the nonlinear field dynamics. 

The Euler-Lagrange equations for the Lagrangian \eqref{lagrangiantruncated} read
\begin{equation}
    \ddot{A}^{(N)}_n+(n^2+\lambda)A^{(N)}_n=\sgn(\lambda)\sum_{mpq}D_{nmpq}A^{(N)}_mA^{(N)}_pA^{(N)}_q,\label{truncatedsystem}
\end{equation}
where $1\leq n\leq N$. Note also that the Hamiltonian corresponding to \eqref{lagrangiantruncated} is conserved by means of Noether's Theorem. In what follows, we shall study two aspects of the system \eqref{truncatedsystem}: the relation between its stationary solutions and the stationary solutions of the KG equation found in Sec.~\ref{stationaryKG}, and how it can be used to approximate the solutions of the Cauchy problem for the nonlinear KG equation.

\subsection{Stationary solutions of the truncated system}

The stationary solutions of the system \eqref{truncatedsystem} are precisely the  critical points of the potential energy $U^{(N)}$. In this section we show how the stationary solutions of the truncated system approximate the exact solutions for the KG equation found in Section \ref{stationaryKG}. In order to keep the consistence of our notation, let us denote by $\mathcal{A}^{(N)}_{i,n}$ the stationary solutions for the truncated system \eqref{truncatedsystem} for negative $\lambda$ and by $\mathcal{C}^{(N)}_{i,n}$ the stationary solutions for positive $\lambda$. Here, the index $i$ labels the distinct critical points according to their energy.  Tables \ref{tabAnN5lambdamenos10} and \ref{tabAnN10lambdamenos10} show the first three stationary points and the corresponding energies of the mechanical systems with $N=5$ and $N=10$ for $\lambda=-10$.
\begin{table}[H]
\centering
\begin{tabular}{c| S S S }
\diagbox[width=2.8em,height=1.5em]{ $n$}{$i$} 
  & \mbox{1}
  &  \mbox{2}
  & \mbox{3} \\
\toprule
\hline
1 & 2.62232   & 0  & 0\\
2 & 0         & 2  & 0   \\
3 & 0.486721  & 0  & 0.816497\\
4 & 0         & 0  & 0\\
5 & 0.11245   & 0  & 0 \\
\hline
$U^{(5)}$ & 9.51594
  &  19.
  & 24.8333\\
\bottomrule
\end{tabular}
\caption{Several values of $\mathcal{A}^{(5)}_{i,n}$ with $5$ precision digits of the first three nontrivial solutions. Here, $\lambda=-10$ and $N=5$.}
\label{tabAnN5lambdamenos10}
\end{table}
\begin{table}[H]
\centering
\begin{tabular}{c| S S S }
\diagbox[width=2.8em,height=1.5em]{ $n$}{$i$} 
  & \mbox{1}
  &  \mbox{2}
  & \mbox{3} \\
\toprule
\hline
1 & 2.62563   & 0  & 0\\
2 & 0         & 2.05109  & 0   \\
3 & 0.493384  & 0  & 0.818358\\
4 & 0         & 0  & 0\\
5 & 0.119297   & 0  & 0 \\
6 & 0   & 0.11261  & 0 \\
7 & 0.0291501   & 0  & 0 \\
8 & 0   & 0  & 0 \\
9 & 0.00703549   & 0  & 0.00375338 \\
10 & 0  & 0.00654  & 0 \\
\hline
$U^{(10)}$ & 9.49029
  &  18.77295
  & 24.8328\\
\bottomrule
\end{tabular}
\caption{Several values of $\mathcal{A}^{(10)}_{i,n}$ with $5$ precision digits of the first three nontrivial solutions. Here, $\lambda=-10$ and $N=10$.}
\label{tabAnN10lambdamenos10}
\end{table}
Comparison between tables \ref{tab1-An}, \ref{tabAnN5lambdamenos10}, and \ref{tabAnN10lambdamenos10} reveals that the truncated system is capable finding the energy stored in the low energy stationary solutions, whose Fourier coefficients are found with good precision already for small values of $N$. The same is also observed for positive $\lambda$, as shown in tables \ref{tab1-Cn}, \ref{tabCnN5}, and \ref{tabCnN10}. 

\begin{figure}[H]
    \centering
    \includegraphics[width=1\linewidth]{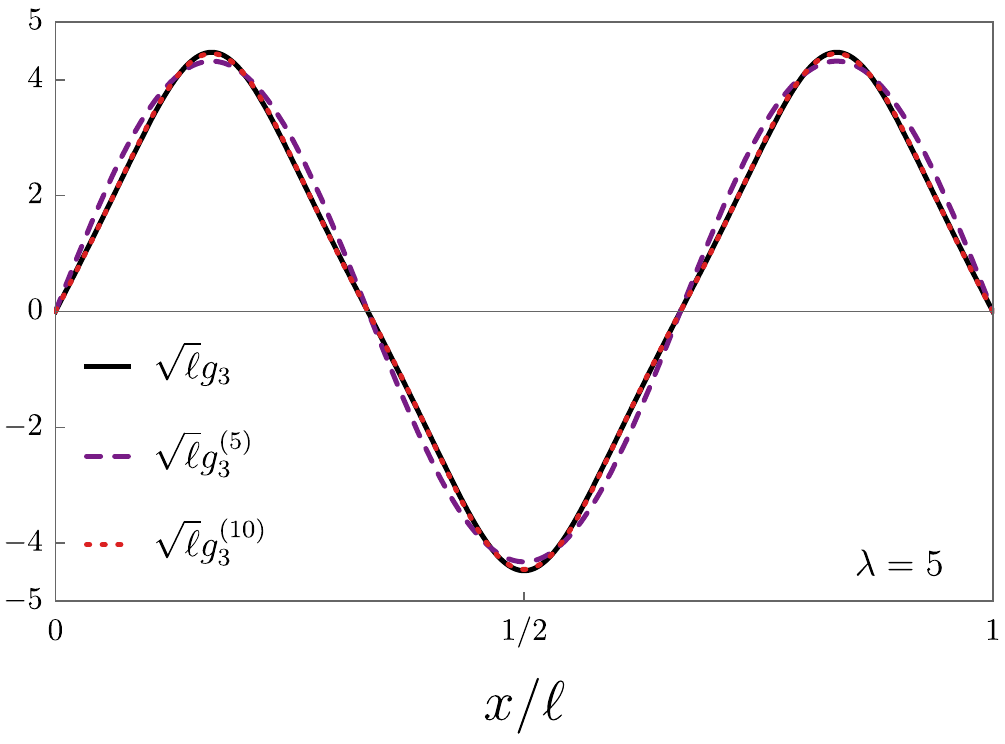}
    \caption{Approximations to the function $g_3$ for $\lambda=5$ via $g^{(5)}_3$ and $g^{(10)}_3$. Notice that the mechanical system with $N=10$ particles already provides a good approximation for the stationary solution $g_3$.}
    \label{figlambda5approx}
\end{figure}

The stationary solutions of the truncated system are also useful to approximate the exact solutions found in Sec.~\ref{stationaryKG}. In fact, we denote by
\begin{align}
    f_i^{(N)}(x)&=\sum_{n=1}^{N}\mathcal{A}^{(N)}_{i,n}\phi_n(x),\label{fitruncada}\\
    g_i^{(N)}(x)&=\sum_{n=1}^{N}\mathcal{C}^{(N)}_{i,n}\phi_n(x),\label{gitruncada}
\end{align}
the approximations for the exact solutions $f_i$, $g_i$, respectively. Figure \ref{figlambda5approx} shows an example of such approximation.
\begin{table}[H]
\centering
\begin{tabular}{c| S S S }
\diagbox[width=2.8em,height=1.5em]{ $n$}{$i$} 
  & \mbox{1}
  &  \mbox{2}
  & \mbox{3} \\
\toprule
\hline
1 & 1.61254   & 0        & 0\\
2 & 0         & 2.44949  & 0   \\
3 & -0.483399 & 0        & 3.05505\\
4 & 0         & 0        & 0\\
5 & 0.114098  & 0        & 0 \\
\hline
$U^{(5)}$ & -1.43409  &  7.25
  & 26.4167\\
\bottomrule
\end{tabular}
\caption{Several values of $\mathcal{C}^{(5)}_{i,n}$ with $5$ precision digits of the first three nontrivial solutions. Here, $\lambda=5$ and $N=5$.}
\label{tabCnN5}
\end{table}
\begin{table}[H]
\centering
\begin{tabular}{c| S S S }
\diagbox[width=2.8em,height=1.5em]{ $n$}{$i$} 
  & \mbox{1}
  &  \mbox{2}
  & \mbox{3} \\
\toprule
\hline
1 & 1.598020496 & 0       &  0\\
2 & 0           & 2.297946 & 0   \\
3 & -0.4889086643& 0       & 2.93684516\\
4 & 0           & 0       & 0\\
5 &0.12326982589& 0       & 0 \\
6 & 0           & -0.2516720       & 0 \\
7 &-0.03055798  & 0       & 0 \\
8 & 0           & 0       & 0 \\
9 & 0.0074530   & 0       &-0.2108829 \\
10 & 0          & 0.02486054       & 0 \\
\hline
$U^{(10)}$& -1.45511853577
  &  6.296703
  & 24.8938481 \\
\bottomrule
\end{tabular}
\caption{Several values of $\mathcal{C}^{(10)}_{i,n}$ with $5$ precision digits of the first three nontrivial solutions. Here, $\lambda=5$ and $N=10$.}
\label{tabCnN10}
\end{table}

\subsection{The Cauchy problem for the truncated system}

In this section we investigate how the finite mechanical system \eqref{truncatedsystem} can be used to approximate the infinite system \eqref{infsystem}. Recall that solving the latter is, by construction, equivalent to solving the nonlinear KG equation, because initial conditions for \eqref{infsystem}, $A_n|_{\tau=0}$, $\dot{A}_n|_{\tau=0}$ for all $n$ are in one-to-one correspondence with initial conditions for $\phi=\varphi/a$ via Eq.~\eqref{fourierexpansion}. Note, also, that $|A_n|\rightarrow0$ when $n\rightarrow\infty$ for any well-behaved field configuration $\phi$, and thus it is reasonable to expect that there exist a positive integer $N$ such that $A_n$, for $n<N$, encapsulate most of the physical content of the field $\phi$. Using this property as motivation, we define 
\begin{equation}
    \phi^{(N)}(t,x)=\sum_{n=1}^{N}A^{(N)}_n(t)\phi_{n}(x),
\end{equation}
with $A^{(N)}_n(t)$ being the solutions to the truncated system \eqref{truncatedsystem}, as a potential approximation to the exact field configuration $\phi$. Furthermore, the resulting error of using this approximation can be studied as follows. By writing the KG equation in terms of $\phi$ as
\begin{equation}
    \ddot{\phi}-\frac{\ell^2}{\pi^2}\partial_x^2\phi-\mbox{sgn}(\lambda)(\ell\phi^2-|\lambda|)\phi=0,
\end{equation}
we define the (local) error as
\begin{align}
    &R^{(N)}=\nonumber\\
    &\sqrt{\ell}\left| \ddot{\phi}^{(N)}-\frac{\ell^2\partial_x^2\phi^{(N)}}{\pi^2}-\mbox{sgn}(\lambda)\left[\ell\left(\phi^{(N)}\right)^2-|\lambda|\right]\phi^{(N)}\right|,\nonumber
    \end{align}
which, in view of Eq.~\eqref{truncatedsystem}, can be written as
    \begin{align}
    &R^{(N)}=\nonumber\\
    &\sqrt{\ell}\left|\sum_{nmp}A^{(N)}_nA^{(N)}_mA^{(N)}_p\left[\sum_{q=1}^{N}D_{nmpq}\phi_{q}-\ell\phi_{n}\phi_{m}\phi_{p}\right]\right|.\label{errorsimple}
\end{align}
We provide numerical simulations of the above formula in what follows.
%
\begin{figure}
    \centering
    \includegraphics[width=1\linewidth]{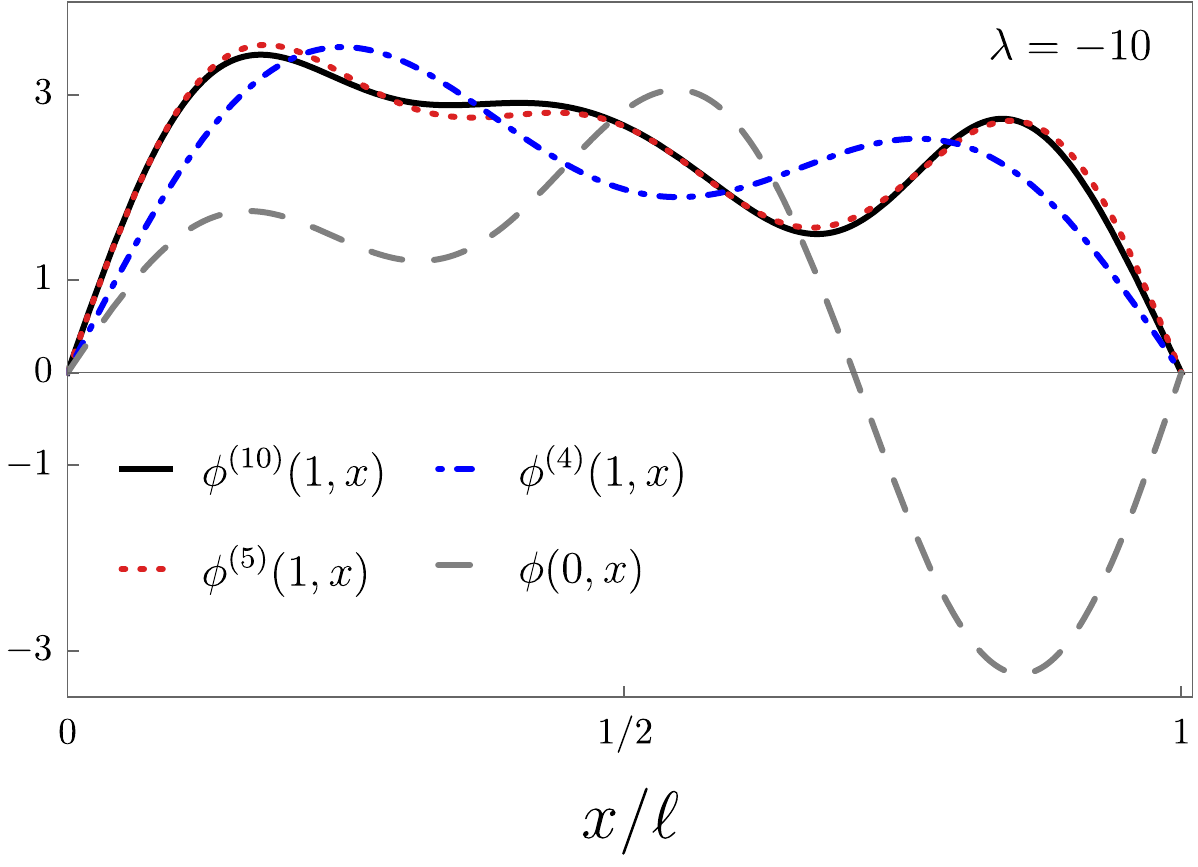}
    \caption{Simulation of $\phi^{(N)}$ for $N=4,5$, and $10$. The initial condition is set as $\phi^{(N)}(0,x)=\phi(0,x)$, where $\phi(0,x)$ is the long-dashed curve in the plot. Also, $\partial_\tau\phi^{(N)}|_{\tau=0}=0$. This initial condition is such that $A_1=A_2=A_4=-A_3=1$, $A_n=0$ for $n>4$, and $\dot{A}_n(0)=0$ for all $n$. Note that although there is a considerable difference between $\phi^{(4)}$ and $\phi^{(5)}$, when we consider $N=10$, for the parameters used in the simulation, the difference between $\phi^{(5)}$ and $\phi^{(10)}$ is smaller. }
    \label{figsimu1}
\end{figure}

\begin{figure}
    \centering
    \includegraphics[width=1\linewidth]{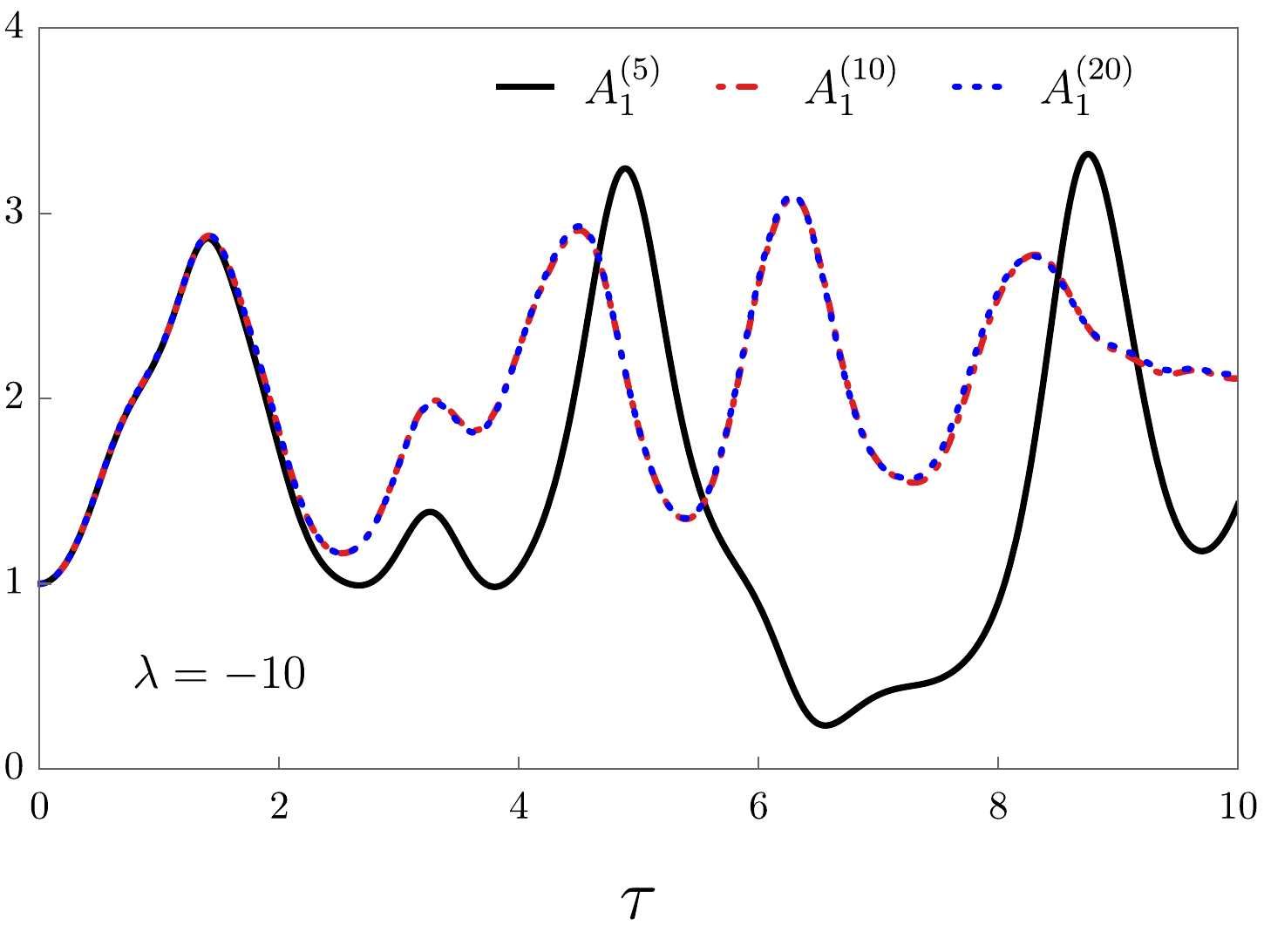}
    \caption{The ``positions'' of the particles $A^{(5)}_1$, $A^{(10)}_1$ and $A^{(20)}_1$  as function of $\tau$. Note that initially they evolve in a similar fashion. However, as time passes, the influence of the extra particles in the systems with $N=10$ (in comparison to $N=5$) leads to a strong deviation between the dynamics of $A_{1}^{(5)}$ and that of $A_{1}^{(10)}$. Note, however, that for the time period adopted in the plot, $A_{1}^{(10)}$ and $A_{1}^{(20)}$ present similar evolution, which is a numerical evidence of the convergence of the method.}
    \label{figsimuA1}
\end{figure}

Figure \ref{figsimu1} depicts the value of $\phi^{(N)}$ for $N=4,5$, and $10$, at $\tau=1$. The parameters are such that $\phi^{(N)}(0,x)=\phi(0,x)$, and $A_1=A_2=A_4=-A_3=1$, $A_n=0$ for $n>4$, with $\dot{A}^{(N)}_n(0)=0$ for all $n$. This means that only the first four particles of the finite mechanical systems are initially excited in this simulation. Note that as we increase the cutoff value from $N=5$ to $N=10$, the change observed on the overall behavior of $\phi^{(N)}$ is small in comparison to the observed change when we increase from $N=4$ to $N=5$, which is indicative of convergence. However, we stress that we set $\tau=1$ in order to compare the simulations, and, in general case, the rate of convergence is dependent on the initial conditions and the evolution time. Figure \ref{figsimuA1} shows how the convergence occurs by plotting $A_{1}^{(5)}$ and $A_{1}^{(10)}$ for $\lambda=-10$ and the same initial conditions as in Fig.~\ref{figsimu1}. Notice that $A_{1}^{(5)}$ and $A_{1}^{(10)}$ present a similar evolution initially and deviate considerably after $\tau=2$. Nevertheless, Fig.~\ref{figsimuA1} also shows $A_{1}^{(20)}$, which is reasonably close to $A_{1}^{(10)}$ within the time window $0<\tau<10$, and this feature is also observed when we compare $A_{i}^{(10)}$ and $A_{i}^{(20)}$, for $2\leq i\leq10$. This is a numerical evidence for the convergence of the method, and illustrates how the finite mechanical systems can be used to study the evolution of the nonlinear KG equation when initial conditions involve excitations of only a few particles.

Therefore, this semianalytical method provides a good approximation for the dynamics of the nonlinear KG equation which depends on the associated initial conditions and time scales. The error generated by the approximation can be studied using Eq.~\eqref{errorsimple}. We also define the total error as $\bar{R}^{(N)}=(1/\ell)\int_0^\ell\d xR^{(N)}$, and show
\begin{figure}
    \centering
    \includegraphics[width=1\linewidth]{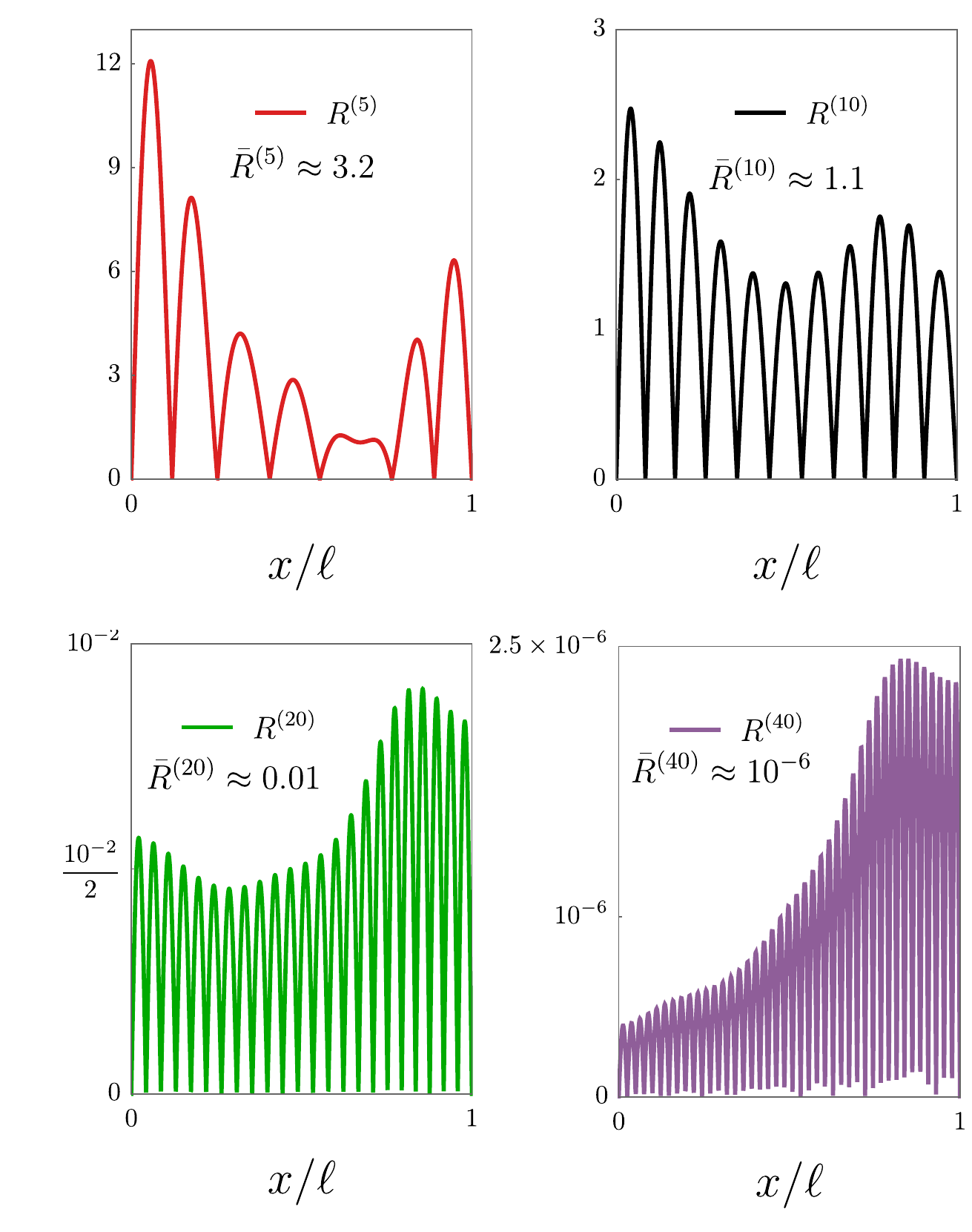}
    \caption{The local and total errors of several simulations. The initial conditions are the same as in Fig.~\ref{figsimu1}, and here $\tau=1$. Notice that the mechanical systems with $N=5$ and $10$ particles cannot approximate the dynamics of the nonlinear field, whereas the system with $N=40$ particles provides a reasonable approximation for the involved time scale and initial conditions.}
    \label{figsimu2}
\end{figure}
in Fig.~\ref{figsimu2} examples of the errors generated by the Galerkin method. The initial conditions are the same as the one used in Fig.~\ref{figsimu1}. Notice that even though in Fig.~\ref{figsimu1} the profiles of $\phi^{(5)}$ and $\phi^{(10)}$ are similar, which suggests convergence, they do not offer reasonable approximations to the KG equation at $\tau=1$, as shown in Fig.~\ref{figsimu2} top right and left panels. The bottom right and left panels depict the local (and total) errors for $N=20$ and $40$, showcasing how the convergence occurs.

\begin{figure}
    \centering
    \includegraphics[width=1\linewidth]{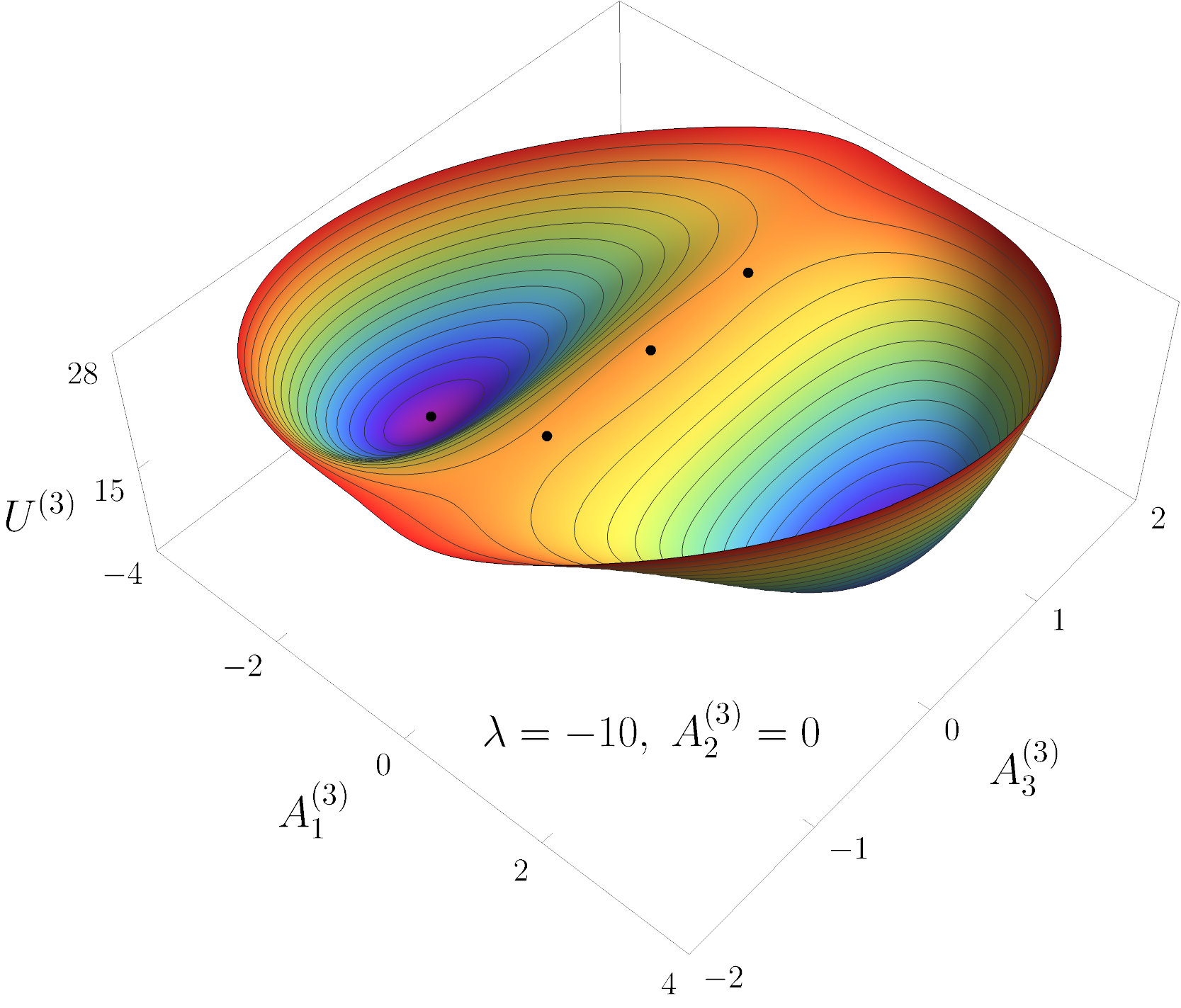}
    \caption{Potential energy $U^{(3)}$ with $A^{(3)}_2\equiv0$ for the case $\lambda=-10$. The black dots indicate the coordinates of the (critical points) stationary solutions of the mechanical system. The plot shows two regions with local minima and all solutions are bounded.}
    \label{figU1}
\end{figure}

\begin{figure}
    \centering
    \includegraphics[width=1\linewidth]{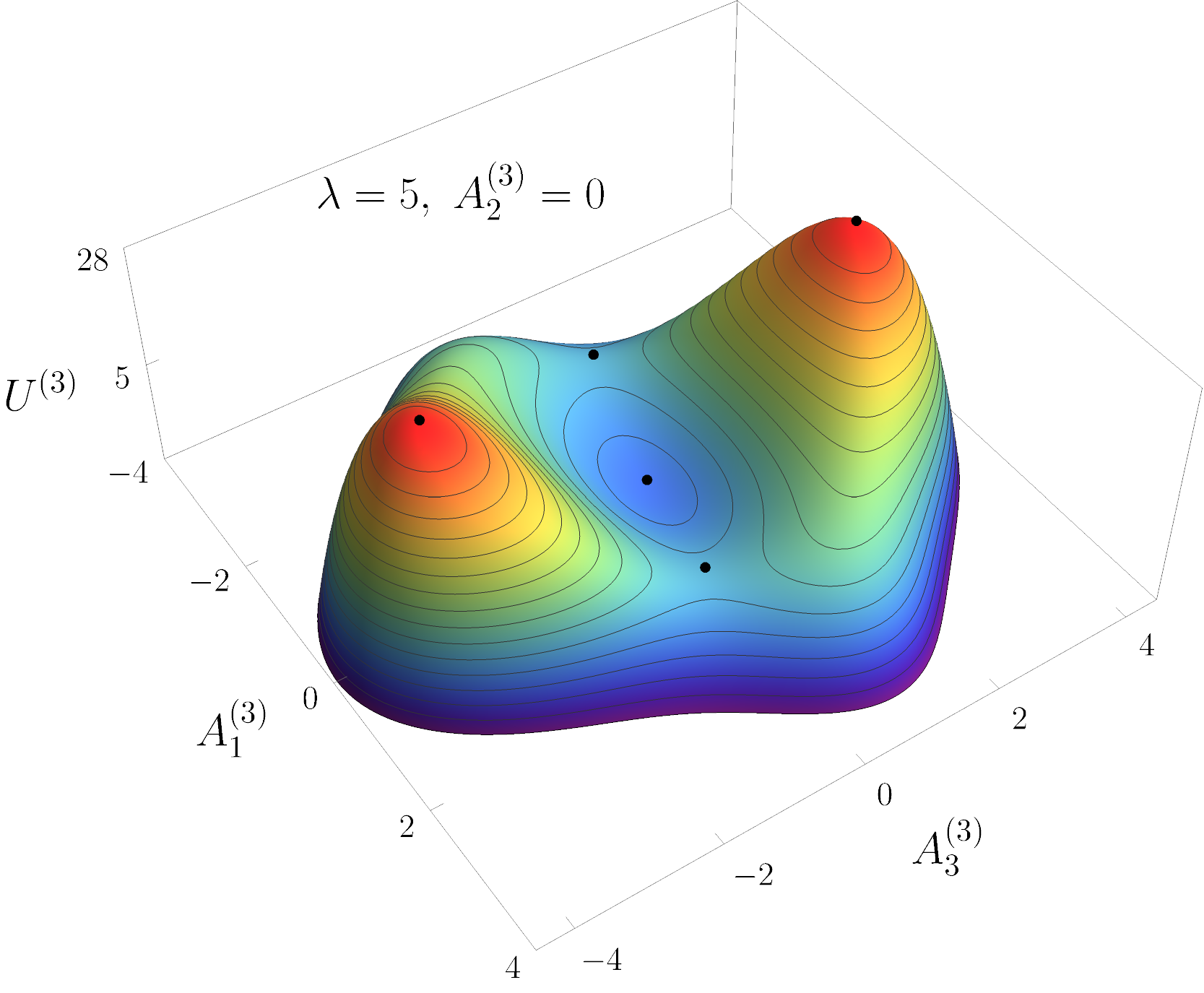}
    \caption{Potential energy $U^{(3)}$ with $A^{(3)}_2\equiv0$ for the case $\lambda=5$. The black dots indicate the coordinates of the (critical points) stationary solutions of the mechanical system. In comparison to the case $\lambda=-10$, the plot shows a richer profile. Of the $5$ critical points, only the one with $A^{(3)}_1=A^{(3)}_3=0$ corresponds to a stable solution (local minimum).}
    \label{figU2}
\end{figure}

In order to gain a deeper insight on some of the physical properties of the truncated mechanical systems, let us consider the case $N=3$. The system \eqref{truncatedsystem} in this case becomes
\begin{align}
    &\ddot{A}_1^{(3)}+(1+\lambda)A_1^{(3)}=\mbox{sgn}(\lambda)\Bigg[\frac{3}{2}A_1^{(3)3}\nonumber\\
    &+3A_1^{(3)}(A_2^{(3)2}+A_3^{(3)2})+\frac{3}{2}A_3^{(3)}(A_2^{(3)2}-A_1^{(3)2})\Bigg],\label{eq1}\\
    &\ddot{A}_2^{(3)}+(4+\lambda)A_2^{(3)}=\mbox{sgn}(\lambda)\Bigg[\frac{3}{2}A_2^{(3)3}\nonumber\\
    &+3A_2^{(3)}(A_1^{(3)2}+A_3^{(3)2})+3A_3^{(3)}A_2^{(3)}A_1^{(3)}\Bigg],\label{eq2}\\
    &\ddot{A}_3^{(3)}+(9+\lambda)A_3^{(3)}=\mbox{sgn}(\lambda)\Bigg[\frac{3}{2}A_3^{(3)3}\nonumber\\
    &+3A_3^{(3)}(A_1^{(3)2}+A_2^{(3)2})+\frac{1}{2}A_1^{(3)}(3A_2^{(3)2}-A_1^{(3)2})\Bigg].\label{eq3}
\end{align}
Notice that although the three particles $A_1^{(3)}, A_2^{(3)}$, and $A_3^{(3)}$ are coupled, there are families of solutions in which some of the particles remain at rest at all times.  In fact, solutions exist for $A^{(3)}_2=0$ and  $A^{(3)}_1,A^{(3)}_3\neq0$, or $A^{(3)}_2\neq0$ and  $A^{(3)}_1,A^{(3)}_3=0$, or $A^{(3)}_3\neq0$ and  $A^{(3)}_1,A^{(3)}_2=0$. Let us study qualitatively the solutions of Eqs.~\eqref{eq1}, \eqref{eq2}, and \eqref{eq3} for $A^{(3)}_2=0$. In this case, we can view $A^{(3)}_1(t)$ and $A^{(3)}_3(t)$ as the coordinates of a single particle of unit mass allowed to move in the plane, with potential energy $U^{(3)}$. We present in Figs.~\ref{figU1} and \ref{figU2} the potential energy when $\lambda=-10$ and $\lambda=5$, respectively. 

Stemming from the fact that the potential \eqref{potentialV} leads to a bounded Hamiltonian when $\beta>0$, i.e., $\lambda<0$, we see from Fig.~\ref{figU1} that all admissible particle trajectories remain bounded. Also, note that the critical points at the local minima have coordinates (up to an overall sign) coinciding with $\mathcal{A}^{(3)}_{1,n}$, which is an element of the sequence $\mathcal{A}^{(N)}_{1,n}$ as function of the cutoff $N$  [see Eq.~\eqref{fitruncada}]. We refer to Tables  \ref{tabAnN5lambdamenos10}, \ref{tabAnN10lambdamenos10} and the discussion therein for the values of $\mathcal{A}^{(5)}_{1,n}$ and $\mathcal{A}^{(10)}_{1,n}$. Note that this critical point occurs at a local minimum and thus perturbations of this stationary solution remain small. 
Similarly, the critical points other than the trivial one correspond to the sequence $\mathcal{A}^{(N)}_{3,n}$ [cf. Tables  \ref{tabAnN5lambdamenos10}, \ref{tabAnN10lambdamenos10}], and these give rise to unstable solutions.   

Now in sharp distinction to the $\lambda<0$ ($\beta>0$) case, the potential $U^{(3)}$ leads to interesting behavior when $\lambda=5$ because the potential Eq.~\eqref{potentialV} is not bounded from below, as shown in  Fig.~\ref{figU2}. Notice that in this case the critical point at the origin corresponds to a stable solution, whereas all the others lead to unstable solutions. Also, the critical points at the local maxima occur at the coordinates given by $\mathcal{C}^{(N)}_{3,n}$ for $N=3$, whereas the saddle points are determined by $\mathcal{C}^{(N)}_{1,n}$ for $N=3$. Tables \ref{tabCnN5} and \ref{tabCnN10} present the coefficients $\mathcal{C}^{(5)}_{i,n}$ and $\mathcal{C}^{(10)}_{i,n}$, respectively.


%
%

%
%

%
%

%
%

\section{Final remarks}

\label{finalremarks}

In this work we applied the Galerkin method to generate approximate solutions to the nonlinear KG equation in $1+1$ dimensions. Specifically, we considered real scalar fields spatially constrained to the interval $[0,\ell]$ and subjected to Dirichlet boundary conditions under the influence of Mexican-hat-like potentials. Following the standard procedure of using suitable bases of functions to expand the field variable, a correspondence between the field theories and mechanical systems of infinitely many particles was constructed, and we provided an interpretation of the approximating method in terms of mechanical systems of only a few particles. Among the results, analytical stationary solutions to the nonlinear KG equation were constructed. These are relevant, for instance, if one is interested in studying properties of the equilibrium configurations. Furthermore, we showed how the method of approximating the solutions is capable of capturing properties of both the stationary and time-dependent solutions.  

Before we close this work, a few remarks are in order. First, we note that there is not an universal numerical method that can be used to solve all partial differential equations, and each individual case must be studied individually. For instance, certain finite-difference methods applied to the the Gross-Pitaevskii equation lead to solutions violating conservations laws \cite{BAO2003318}. In this context, the Galerkin method has an interesting feature: if the field equation under study comes from a Lagrangian, then one has control of the conserved quantities in the approximating systems, as they also originate from a Lagrangian. 

A second and important aspect of our analysis relates the convergence rate of the method and the choice of a complete set of functions to expand the field variable. For the real nonlinear KG equation with Dirichlet BC considered in our work, a ``natural'' choice of functions was provided by the linear part of the field operator. These led to approximate solutions that satisfied the BC automatically. However, a different  complete set of functions could be equally adopted. For instance, if the KG equation were subjected to Neumann BC and the same functions $\phi_n$ of Eq.~\eqref{dirichletbf} were used, then the convergence of the method would be affected, because a larger number of mode functions, that vanish on the spatial boundary, would be necessary to expand field configurations that do not vanish at the spatial boundary.

Finally, we recall that correspondences between the stationary solutions of the nonlinear KG equation and the critical points of the truncated mechanical systems were found for the particular system under study. We stress, however, that there is no guarantee that such correspondences will always appear, and the fact that they are present here should be viewed as a feature of the method applied to the KG equation under Dirichlet BC and using as a basis the set $\{\phi_n\}$.

\section*{Acknowledgements} 
C.C.H.R. would like to thank the Funda\c{c}\~ao de Apoio \`a Pesquisa do Distrito
Federal (grant 00193-00002051/2023-14) for supporting this work. L.L.S.R. was supported by Funda\c{c}\~ao de Amparo \`a Pesquisa do Estado de Minas Gerais (grant APQ-01574-24).

\appendix
\bibliography{qgav3.bib}
\end{document}